\documentclass[galaxies,article,submit,moreauthors,pdftex,10pt,a4paper]{mdpi_noline} 
\firstpage{1} 
\articlenumber{x}
\doinum{10.3390/------}
\pubvolume{xx}
\pubyear{2017}
\copyrightyear{2017}
\externaleditor{Academic Editor: name}
\history{Received: date; Accepted: date; Published: date}

\Title{SALT Spectropolarimetry and Self-Consistent SED and Polarization Modeling of 
Blazars\footnote[1]{based on observations made with the Southern African Large Telescope 
(SALT) under programme 2016-2-LSP-001 (PI: D. A. H. Buckley)}}

\orcidauthorONE{0000-0002-8434-5692} 

\Author{Markus B\"ottcher$^{1}$\orcidONE{0000-0002-8434-5692}, Brian van Soelen$^{2}$,
Richard J. Britto$^{2}$\footnote[2]{for the Fermi-LAT collaboration}, David A. H. Buckley$^{3}$, 
Johannes P. Marais$^{2}$ and Hester Schutte$^{1}$}

\AuthorNames{Markus B\"ottcher, Brian van Soelen, Richard J. Britto, David A. H. Buckley,
Johannes P. Marais and Hester Schutte}

\address{%
$^{1}$ \quad Centre for Space Research, North-West University, Potchefstroom, South Africa; 
Markus.Bottcher@nwu.ac.za\\
$^{2}$ \quad Department of Physics, University of the Free State, Bloemfontein, South Africa \\
$^{3}$ \quad South African Astronomical Observatory, Cape Town, South Africa}

\corres{Correspondence: Markus.Bottcher@nwu.ac.za; Tel.: +27-18-299-2418}

\abstract{
We report on recent results from a target-of-opportunity program to
obtain spectropolarimetry observations with the Southern African Large Telescope 
(SALT) on flaring gamma-ray blazars. SALT spectropolarimetry and contemporaneous
multi-wavelength spectral energy distribution (SED) data are being modelled
self-consistently with a leptonic single-zone model. Such modeling provides 
an accurate estimate of the degree of order of the magnetic field in the emission 
region and the thermal contributions (from the host galaxy and the accretion
disk) to the SED, thus putting strong constraints on the physical parameters 
of the gamma-ray emitting region. For the specific case of the $\gamma$-ray 
blazar 4C+01.02, we demonstrate that the combined SED and spectropolarimetry
modeling constrains the mass of the central black hole in this blazar to
$M_{\rm BH} \sim 10^9 \, M_{\odot}$.
}

\keyword{active galaxies --- blazars; gamma-rays; polarization --- optical; spectropolarimetry}

\begin{document}



\section{\label{Intro}Introduction}

Blazars are a class of radio-loud, jet-dominated active galactic nuclei whose jets are closely 
aligned with our line of sight. Due to relativistic Doppler boosting, they are bright and often
rapidly variable sources across the entire electromagnetic spectrum. Their spectral energy 
distribution (SED) is dominated by two broad non-thermal components. The low-frequency component 
(radio through optical -- X-rays) is well understood as synchrotron radiation from relativistic 
electrons, while for the high-energy component (X-rays through $\gamma$-rays) both leptonic and 
hadronic emission mechanisms are possible \citep[e.g.,][]{Boettcher07,Boettcher13,Romero17}. 
In leptonic models, the high-energy emission is produced by Compton 
scattering of soft radiation fields off the same relativistic electrons 
producing the synchrotron emission. A variety of target photon fields
are possible, including the co-spatially produced synchrotron radiation and external photon
fields from the accretion disk, the broad-line region, and/or the infrared-emitting dust
torus around the central engine \citep{Boettcher07,Ghisellini10,Boettcher13}. In the infrared
through UV, also thermal components from the host galaxy and the accretion-disk + dust-torus
system add to the SED. Due to the multitude of plausibly contributing radiation components and
our lack of knowledge of the dominant particle acceleration mechanism, there is significant 
ambiguity concerning the underlying particle distributions and the location of and physical 
conditions within the $\gamma$-ray emission zone \citep[e.g.,][]{Ghisellini10,Boettcher13}. These 
degeneracies can generally not be broken by SED modeling alone. 

An additional aspect of radiation, which provides important information, is polarization. While 
radio polarization is routinely employed to diagnose magnetic-field topologies in the large-scale 
jets of radio-loud AGN (one of the main topics of this conference), the utility of optical
and high-energy (X-ray / $\gamma$-ray) polarimetry as a disgnostic for the physical conditions in 
the high-energy emission region has only recently begun to be considered 
\citep[e.g.,][]{Zhang13,Marscher14,Zhang14,Zhang15}. In view of a multitude of on-going optical 
polarimetric blazar monitoring programs, this provides a promising avenue for progress in
our understanding of the physical conditions in AGN jets. Specifically, the optical emission from
blazars is often dominated by synchrotron emission from the jet, which is well-known to be polarized
at a degree $\Pi_{\rm sy}$ related to the spectral index $p$ of the underlying non-thermal electron 
distribution (resulting in a synchrotron radiation spectral index $\alpha = (p - 1)/2$) and the 
degree of order of the magnetic field $f_{\rm B, order}$ through

\begin{equation}
\Pi_{\rm sy} = f_{\rm B, order} \, {p + 1 \over p + 7/3} = f_{\rm B, order} \, {\alpha + 1 \over \alpha + 5/3}
\label{Pisy}
\end{equation}

Additional contributions to the optical spectrum may arise from the host galaxy and a dust 
torus, which may be dominant towards the red,
and from the accretion disk, which may contribute significantly towards the blue end of the spectrum.
These thermal contributions are expected to be unpolarized and will reveal their presence through a 
decline of the degree of polarization (compared to pure synchrotron emission) in spectropolarimetric 
observations. 

The above considerations motivated us to use the Southern African Large Telescope (SALT --- see Section
\ref{SALT}) for spectropolarimetry of flaring {\it Fermi}-detected $\gamma$-ray blazars in order to constrain
the degree of ordering of the magnetic field in the emission region and the contribution of thermal
radiation components to the optical spectrum. We focus on the specific example of the high-redshift 
($z = 2.1$) blazar 4C+01.02, which underwent a large $\gamma$-ray flare in July 2016. The resulting 
SEDs and spectropolarimetry results were modelled with a leptonic single-zone model that self-consistently 
calculates the SEDs along with the expected optical spectropolarimetry signatures (see Section \ref{Model}). 
Results from this combined fitting procedure for the case of 4C+01.02 are presented and discussed in 
Sections \ref{Results} and \ref{Summary}.

\section{\label{Obs}Observations}

Our observations were conducted in the framework of a SALT Large Program on high-energy transients
(PI: D. Buckley). This program aims at target-of-opportunity (ToO) spectroscopy and spectropolarimetry
observations of various transients, including cataclysmic variables, novae, microlensing events,
X-ray binaries, tidal disruption events, gamma-ray bursts, and flaring blazars. In the case of 
blazars, ToO observations are triggered by flaring activity detected by {\it Fermi}-LAT and, when 
possible, complemented by optical photometry using the Las Cumbres Observatory (LCO) network and 
X-ray observations with {\it Swift}. Within 
the first year of this program (April 2016 -- May 2017), 14 ToO spectropolarimetry observations 
targeting 7 different blazars (5 Flat Spectrum Radio Quasars [FSRQs] and 2 BL Lac objects) were 
conducted, revealing generally frequency-dependent polarization degrees 
in the range $\sim 5$ -- $25$~\%. In the following we focus on the particularly interesting case 
of the FSRQ 4C +01.02 (PKS B0106+013; $z = 2.1$), which underwent a large $\gamma$-ray flare in 
July 2016.

\subsection{\label{Fermi}Fermi-LAT}

\begin{figure}[t]
\centering
\includegraphics[width=9cm]{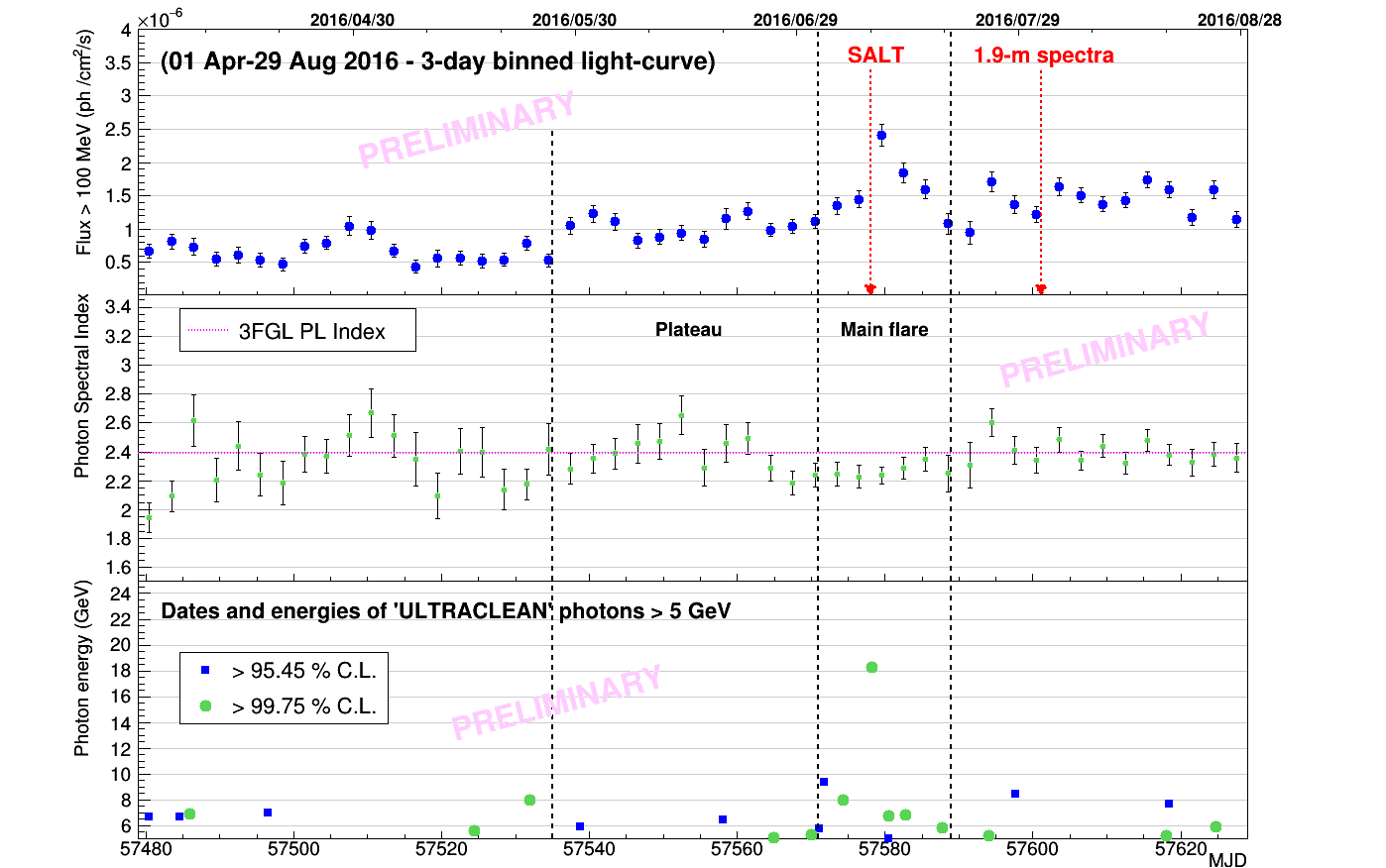}
\caption{{\it Top panel:} 3-day binned {\it Fermi}-LAT light curve at $E > 100$~MeV of 4C+01.02 for the
period 2016 April -- June. {\it Middle panel:} {\it Fermi}-LAT spectral index light curve for a power-law
fit to the 3-day binned spectra. {\it Bottom panel:} Arrival times and energies of high-energy ($E > 5$~GeV) 
photons.}
\label{FermiLC}
\end{figure}

The {\it Fermi} Large Area Telescope (LAT), is observing the whole sky every 3 hr, which enables 
an almost continuous monitoring of the flux variability of bright sources. {\it Fermi}-LAT is 
sensitive to photons from 20~MeV to $>$ 300 GeV \citep{Atwood09}. We analysed data from 4C+01.02 
between 100~MeV and 300~GeV during 
its high state in 2016, using the {\it Pass 8} data representation \citep{Atwood13} and 
the {\it Fermi} Science Tools version v10r0p5.\footnote{http://fermi.gsfc.nasa.gov/ssc/data/analysis/}
We applied the following standard analysis cuts: radius of the \emph{Region of interest} 
(ROI)=$15^\circ$; Source region=ROI$+10^\circ$; {\sc source} class; front + back event type; zenith 
angle $< 90^\circ$; {\sc DATA\_QUAL=1, LAT\_CONFIG=1}; Diffuse emission: \texttt{gll\_iem\_v06.fits} 
(Galactic) and \texttt{iso\_P8R2\_SOURCE\_V6\_v06.txt} (extragalactic) templates. In the light-curve 
processing, we used the unbinned likelihood gtlike/pyLikelihood tool, and modeled the source of 
interest by a single power law spectrum of photon index $\Gamma$. In the SED processing, we used 
the binned likelihood analysis from the same tool, within the {\it Enrico} Python package 
\cite{Sanchez13}, for a preliminary SED construction. Fig. \ref{FermiLC} shows the light curve of 
4C+01.02 during 11 April -- 29 August, 2016. We labeled the July 2 -- 20 period as ``Main flare" 
as it contains the highest peak in flux for the considered high state, which also was the highest 
flux ever detected with {\it Fermi}-LAT for this source to date 
($\simeq 2.8 10^{-6} \pm 0.3$ ph cm$^{-2}$ s$^{-1}$ $>$ 100 MeV on 10 July [MJD 57579], daily 
average). The SED of the {\it main flare} is plotted in Fig. \ref{fits} along with multiwavelength 
data.

\subsection{\label{SALT}SALT}

The Southern African Large Telescope (SALT) is a 10-m class telescope, located at the South African 
Astronomical Observatory (SAAO), near Sutherland, South Africa \citep{buckley06}. One of its main 
instruments is the {\it Robert Stobie Spectrograph (RSS)}, located at the telescope's prime focus. 
The RSS is capable of performing spectroscopy in various modes, including long-slit,
multi-object slit, polarimetry, and Fabry-P\'erot, and is sensitive to 
wavelengths from $\sim$3200 \AA\ to $\ge$9000 \AA\ \cite{RSS1,RSS2}. The results 
presented in Fig. 2 were obtained using RSS in the spectropolarimetry ``{\sc LINEAR}'' mode \cite{RSS_pol} 
on 2016 July 9 and November 27, 28 and 29. Each spectrum, along with the polarization degree 
and polarization angle as a function of wavelength, was obtained through four exposures of 600 s, one for 
each of the four orientations of the half-wave plates (0$^\circ$, 45$^\circ$, 22.5$^\circ$ and 67.5$^\circ$ 
respectively). Data reduction was performed using the {\sc polsalt} reduction 
pipeline.\footnote{https://github.com/saltastro/polsalt}

\begin{figure}[t]
\centering
\hbox{\includegraphics[width=8cm]{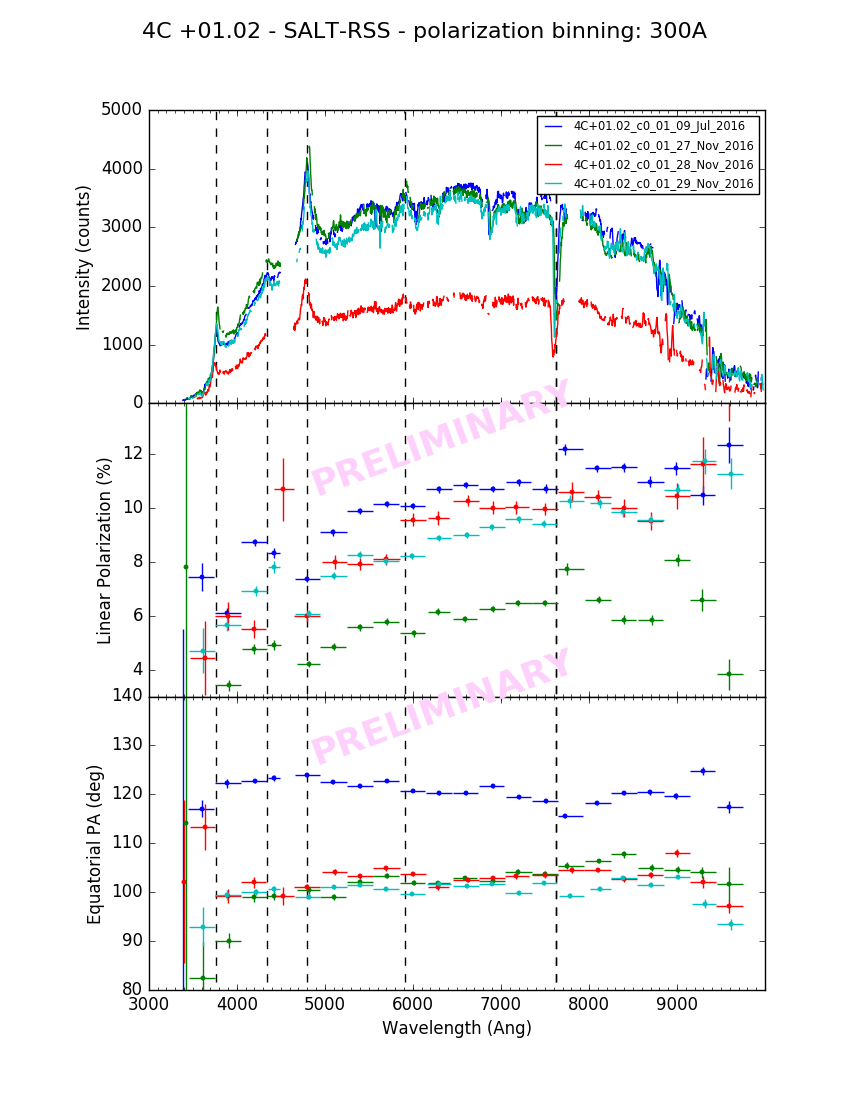} \qquad \vbox{\hsize=0.4\hsize {\bf Figure 2.}
{\it Top panel:} SALT RSS count spectra of 4C+01.02 during four observing 
  windows, including the 2016 July 9 flare indicated in Fig. \ref{FermiLC} (blue). Dashed vertical
  lines indicate the positions of the following lines: Ly$\alpha$ 1216 \AA, Si {\sc IV} 1400 \AA,
  C {\sc IV} 1549 \AA, C {\sc III]} 1909 \AA\ and a telluric absorption line. {\it Middle
panel:} Linear polarization degree as function of wavelength from SALT spectropolarimetry for the
four SALT observing windows, as in the top panel. {\it Bottom panel:} Polarization angle as a
function of wavelength for the four SALT observations. \bigskip\bigskip\bigskip\bigskip\bigskip}}
\label{SALTspectropol}
\end{figure}

\subsection{\label{LCO}LCO}

The Las Cumbres Observatory (LCO) is a global network of 18 telescopes located at eight different 
locations \cite{brown13}. Observations were undertaken with the 1-m class telescopes using the SBIG CCD. 
Data reduction followed the standard procedures using the {\sc iraf/noao} packages. Differential 
photometry was performed using four nearby comparison stars whose magnitudes were taken from the 
NOMAD catalog \cite{zacharias04}. The photometric data points included in Fig. \ref{fits} were
taken on 2016, August 2.

\section{\label{Model}SED and Spectropolarimetry Modeling}

The SEDs of blazars are often successfully modelled with simple single-zone leptonic
radiation models. In this paper, we adopt the time-independent leptonic blazar emission 
model described in detail in \cite{Boettcher13}. The code evaluates the non-thermal 
synchrotron + Compton radiation spectrum based on an equilibrium solution for
the electron spectrum, assuming a rapid acceleration mechanism that provides a power-law
injection spectrum with index $q$ between cut-off energies $\gamma_{\rm min}$ and $\gamma_{\rm max}$
that is balanced by self-consistent radiative losses and escape on a time scale $t_{\rm esc} 
= \eta_{\rm esc} \, R/c$ where $R$ is the size of the spherical emission region. The non-thermal 
jet emission, direct accretion-disk emission, plus the dominant Ly$\alpha$ and C IV emission 
lines from the BLR (see Fig. 2, top panel), are added up to yield the total SED. 

The degree of polarization of the synchrotron emission is evaluated based on the local 
synchrotron spectral index $\alpha$ through Equ. (\ref{Pisy}). The frequency-dependent 
degree of polarization of the total observed optical spectrum is then calculated assuming
that the direct accretion-disk ($F_{\rm AD}$) + BLR line ($F_{\rm line}$) emissions are 
unpolarized, so that

\begin{equation}
\Pi (\nu) = {\Pi_{\rm sy} (\nu) \, F_{\rm sy} (\nu) \over F_{\rm sy} (\nu) + F_{\rm AD} (\nu)
+ F_{\rm line} (\nu)}
\label{Pitotal}
\end{equation}
Our fitting procedure consists of adjusting parameters of the model to obtain simultaneous 
agreement with the SED and spectropolarimetry data of 4C+01.02 during the $\gamma$-ray
flare of July 2016. Since only the optical and $\gamma$-ray data are contemporaneous, we 
include the archival {\it Swift}-XRT X-ray spectrum from summed observations in 2007 and 
2008 \citep{Ghisellini11} as a guide for our spectral fit since they also correspond to a 
moderately high $\gamma$-ray flux state of the source. 

\setcounter{figure}{2}

\begin{figure}[t]
\centering
{\includegraphics[width=6.5cm]{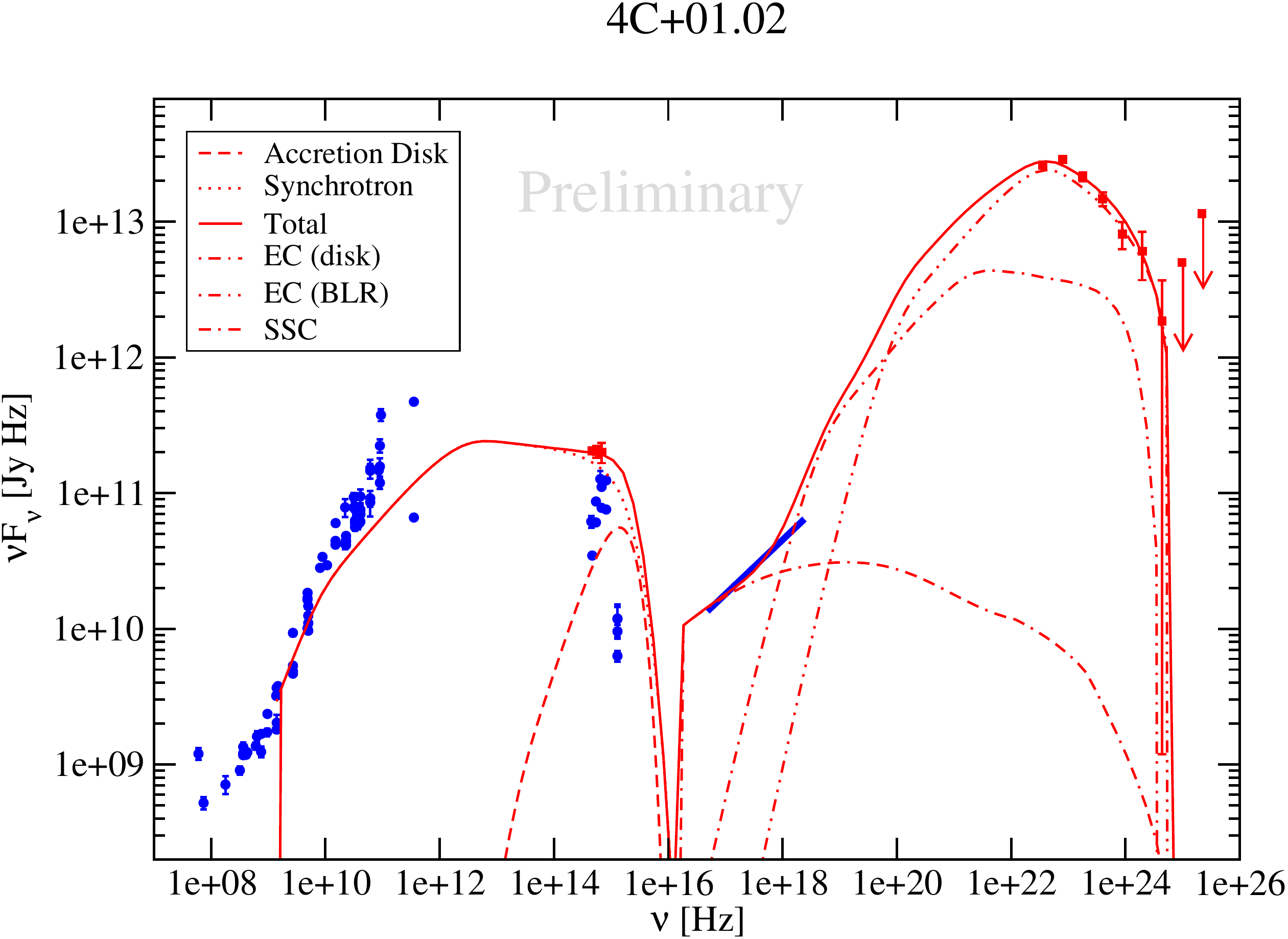} \qquad \includegraphics[width=6.5cm]{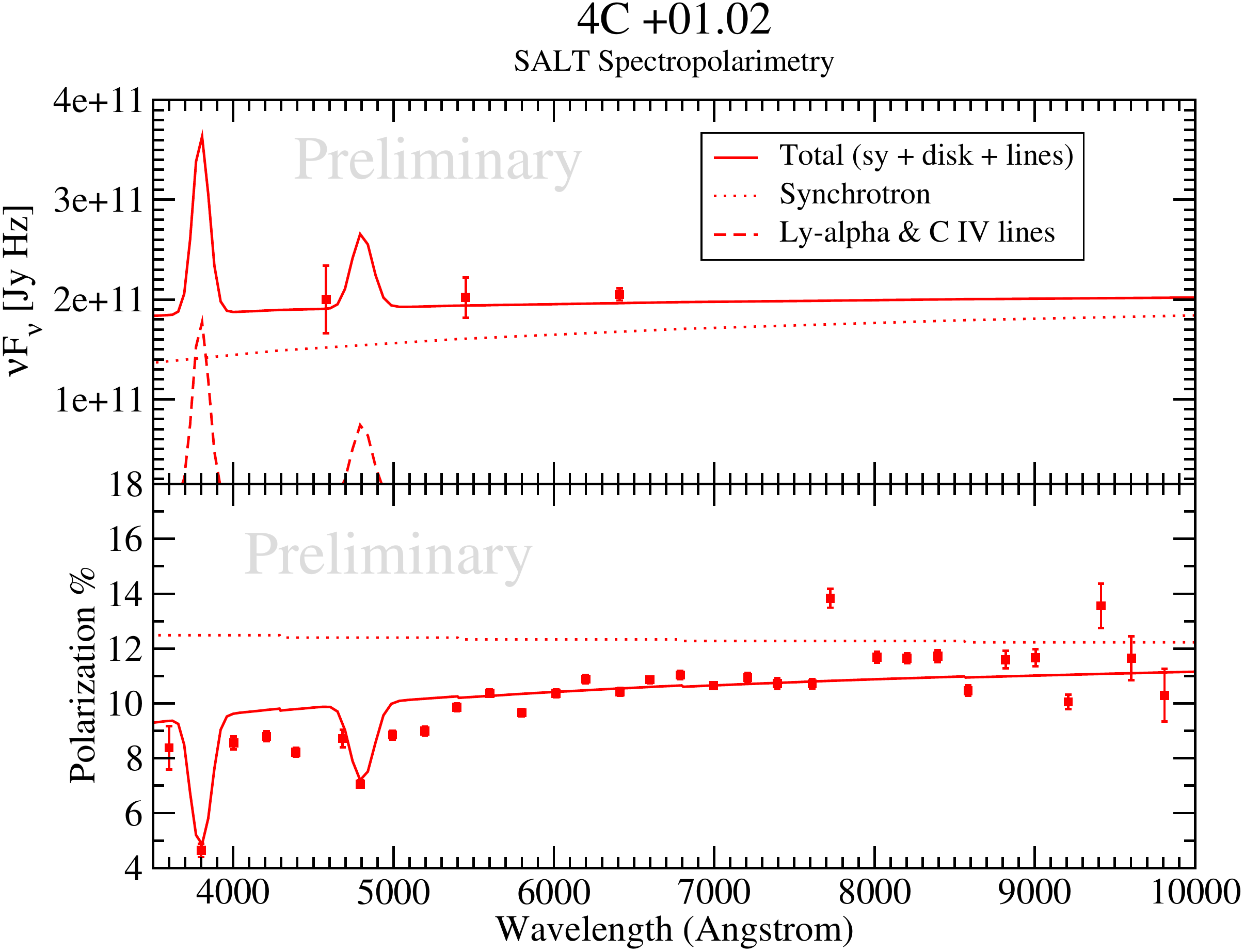}}
\caption{{\it Left panel:} Contemporaneous optical + $\gamma$-ray SED (red) of 4C+01.02, along 
with archival data (from NED$^1$ -- blue), and our SED model (solid red). {\it Right panel:} {\it Top:} 
Optical spectrum with the 3 LCO photometry points, with our model (red lines). {\it Bottom:} SALT 
spectropolarimetry (data points), along with our model prediction (solid red line), and the degree 
of polarization of the synchrotron radiation component (red dotted).}
\label{fits}
\end{figure}

\footnotetext[1]{\tt https://ned.ipac.caltech.edu/}

\section{\label{Results}Results}

Fig. \ref{fits} shows the result of our combined SED + spectropolarimetry fit to 4C+01.02 in 
July 2016. The parameters used for this fit are listed in Table \ref{parameters}. 

\begin{table}[t]
\caption{Model parameters.}
\centering
\begin{tabular}{cc}
\toprule
\textbf{Parameter}	& \textbf{Value} \\
\midrule
Electron injection $\gamma_{\rm min}$ & 230 \\
Electron injection $\gamma_{\rm max}$ & $6 \times 10^3$ \\
Electron injection index $q$ & 2.1 \\
Kinetic power in radiating electrons $L_e$ & $2.5 \times 10^{45}$~erg/s \\
Doppler factor $\delta$ & 35 \\
Magnetic field $B$ & 1 G \\
Magnetic-field order $f_{\rm B, order}$ & 0.16 \\
Energy partition ratio $L_B / L_e$ & 0.45 \\
Escape time-scale parameter $\eta_{\rm esc}$ & 1 \\
Emission-region radius $R$ & $2.2 \times 10^{16}$~cm \\
Emission-region distance from BH & 0.2 pc \\
Black-hole mass $M_{\rm BH}$ & $10^9 \, M_{\odot}$ \\
Accretion-disk luminosity $L_d$ & $1.26 \times 10^{47}$~erg/s = $L_{\rm Edd}$ \\
External radiation field $u_{\rm ext}$ & $5 \times 10^{-3}$~erg/cm$^3$ \\
\bottomrule
\end{tabular}
\label{parameters}
\end{table}

The chosen parameters result in a minimum allowed variability time scale from causality arguments of
$t_{\rm var, min} = R / (\delta \, c) = 18$~hr, consistent with the lack of evidence for intra-day 
variability. 

The peak of the accretion-disk emission is well constrained by the spectropolarimetry results to be 
located near or beyond the blue end of the optical spectrum, requiring a central black-hole
mass of $M_{\rm BH} \leq 10^9 \, M_{\odot}$. On the other hand, its significant contribution to the 
observed spectrum requires a luminosity of the order of the Eddington luminosity for a $10^9 \, M_{\odot}$
black hole. Thus, the mass of the central black hole in 4C+01.02 is well constrained to be $M_{\rm BH} \sim
10^9 \, M_{\rm BH}$.

\section{\label{Summary}Summary and Conclusions}

We have presented results of ToO spectropolarimetry observations using the SALT RSS, focusing on 
the case of the high-redshift FSRQ 4C+01.02 during a large $\gamma$-ray flare in July 2016. The 
contemporaneous SED and spectropolarimetry data were fitted in a consistent way with a single-zone 
leptonic model. This combined fitting constrained the accretion-disk contribution to the SED and,
in particular, the black-hole mass in the center of 4C+01.02 to $M_{\rm BH} \sim 10^9 \, M_{\odot}$. 
This is in tension with the claim by \cite{Ghisellini11} who modelled the entire optical -- UV 
spectrum of the source to be dominated by direct accretion-disk emission, requiring a black-hole 
mass of $M_{\rm BH} = 5 \times 10^9 \, M_{\odot}$. If such an interpretation of the optical -- UV
emission from 4C+01.02 is correct, the degree of polarization in the optical should be close to 
zero. This contradicts our SALT spectropolarimetry results which indicate that even in 
the moderate-activity state in November 2016, the optical polarization is still of the order of
$\sim 5$ -- 10~\%. Our combined SED + spectropolarimetry modeling therefore strongly favours a
synchrotron-dominated optical emission scenario with the accretion-disk contributing only towards
the blue end of the spectrum, requiring a black-hole mass of $M_{\rm BH} \sim 10^9 \, M_{\odot}$.

\acknowledgments{The work of M.B. is supported through the South African Research Chairs Initiative (SARChI)
of the South African Department of Science and Technology (DST) and National Research 
Foundation\footnote{Any opinion,
finding and conclusion or recommendation expressed in this material is that of the authors, and the NRF does 
not accept any liability in this regard.}. The authors acknowledge further
support by the South African DST through the South African Gamma-Ray 
Astronomy Programme. The authors are grateful for the support provided by S. Crawford and K.H. Nordsieck 
with reducing the spectropolarimetry observations. This work makes use of observations from the LCO network.
Some of the observations reported in this paper were obtained with the Southern African Large Telescope (SALT).

The \textit{\emph{Fermi}}-LAT Collaboration acknowledges support for LAT development, operation and data 
analysis from NASA and DOE (United States), CEA/Irfu and IN2P3/CNRS (France), ASI and INFN (Italy), MEXT, 
KEK, and JAXA (Japan), and the K.A.~Wallenberg Foundation, the Swedish Research Council and the National 
Space Board (Sweden). Science analysis support in the operations phase from INAF (Italy) and CNES (France) 
is also gratefully acknowledged.}










\end{document}